\documentclass[12pt]{article}
\usepackage{amssymb,amsmath}
\usepackage{latexsym}
\usepackage{graphicx}
\usepackage{psfrag}

\numberwithin{equation}{section}
\def\coeff#1#2{\relax{\textstyle {#1 \over #2}}\displaystyle}

\def\IR{\mathbb{R}}
\def\ZZ{\mathbb{Z}}

\def\cF{{\cal F}}
\def\cN{{\cal N}}
\def\cP{{\cal P}}

\def\cR{{\cal R}}

\def\cT{{\cal T}}
\newcommand{\be}{\begin{equation}}
\newcommand{\ee}{\end{equation}}
\newcommand{\bea}{\begin{eqnarray}}
\newcommand{\eea}{\end{eqnarray}}

\def\nn{\nonumber \\}

\begin{document}

\begin{titlepage}

\begin{flushright}
IPhT-T09/137
\end{flushright}

\bigskip
\bigskip
\centerline{\Large \bf Supergravity Solutions from Floating Branes}
%\centerline{\Large }
\bigskip
\bigskip
\centerline{{\bf Iosif Bena$^1$, Stefano Giusto$^1$,}}
\centerline{{\bf Cl\'{e}ment Ruef$^{\, 1}$ and Nicholas P. Warner$^2$}}
\bigskip
\centerline{$^1$ Institut de Physique Th\'eorique, }
\centerline{CEA Saclay, 91191 Gif sur Yvette, France}
\bigskip
\centerline{$^2$ Department of Physics and Astronomy}
\centerline{University of Southern California} \centerline{Los
Angeles, CA 90089, USA}
\bigskip
\centerline{{\rm iosif.bena@cea.fr,~stefano.giusto@cea.fr, } }
\centerline{{\rm clement.ruef@cea.fr,~warner@usc.edu} }
\bigskip
\bigskip

\begin{abstract}
  \noindent We solve the equations of motion of five-dimensional
  ungauged supergravity coupled to three $U(1)$ gauge fields using a
  floating-brane Ansatz in which the electric potentials are directly
  related to the gravitational warp factors. We find a new class of
  non-BPS solutions, that can be obtained linearly starting from an
  Euclidean four-dimensional Einstein-Maxwell base. This class -- the
  largest known so far -- reduces to the BPS and almost-BPS solutions
  in certain limits. We solve the equations explicitly when the base
  space is given by the Israel-Wilson metric, and obtain
  solutions describing non-BPS D6 and anti-D6 branes kept in
  equilibrium by flux.  We also examine the action of spectral flow on
  solutions with an Israel-Wilson base and show that it relates these
  solutions to almost-BPS solutions with a Gibbons-Hawking base.

\end{abstract}

\end{titlepage}
%%%%%%%%%%%%%%%%%%%%%%%%%%%%%%%%%%%%%

%\tableofcontents

%%%%%%%%%%%%%%%%%%%%%%%%%%%%%%%%%%%%%
\section{Introduction}
%%%%%%%%%%%%%%%%%%%%%%%%%%%%%%%%%%%%%

The classification of supergravity solutions is an important program
that has yielded an amazing amount of new physics and, in particular,
has greatly enhanced our understanding of the $AdS$-CFT
correspondence, the non-perturbative dynamics of string theory, and
the physics of black holes. In pure gravity, the solutions that have
been classified either contain a horizon, or have enough Killing
symmetries to reduce to a two-dimensional problem that can be solved
using integrability (see, for example, \cite{Belinski:2001ph}).

In supergravity and string theory most of the classification work has
focussed on supersymmetric solutions, and is done essentially by using
Killing spinors or $G$-structures to reduce the second-order supergravity equations of
motion to first-order equations  (see, for example, \cite{g-structures}). It is clearly
important to extend this work to non-supersymmetric solutions, not
only because we would like to better understand non-supersymmetric
physics, but also because we expect (from the dynamics of string
theory probes) to find rather large classes of non-supersymmetric
solutions, with very interesting properties. It is fair to say that
progress in this direction has been rather sporadic, and one of the
main reasons appears to be the absence of a guiding structure that is
neither too restrictive (thus allowing interesting solutions), nor too
lax (so that one can actually solve the equations of
motion).

The broader aim of this paper is to provide a guiding structure for
finding interesting, non-supersymmetric solutions of five dimensional
$U(1)^3$ ungauged supergravity. These solutions uplift to solutions of
eleven-dimensional supergravity on a six-torus, and when compactified
to four-dimensions correspond to solutions of the STU model. In the
Ansatz we use, the warp factors and the electric potentials are
equal and hence the probe M2 branes that have the same charge vector
as the solution feel no force. Therefore, we will call this the
``floating brane'' Ansatz.

This Ansatz naturally incorporates the known BPS
\cite{5dsugra,gutowski-reall,Bena:2004de} and almost-BPS
\cite{Goldstein:2008fq} solutions of five-dimensional ungauged
supergravity\footnote{For a few recent examples of almost-BPS
  solutions see \cite{Bena:2009ev,recent,Bena:2009qv,perz}.}, but, as we
  will see, the equations governing the general floating-brane
  solutions are much more general. The mass of these solutions depends
  linearly on their M2 charges (this comes from the equality of the
  warp factors and electric potential) and thus many of the
  floating-brane solutions will be extremal, but some there are also some interesting (but
  rather restrictive) classes of non-extremal floating-brane solutions
  \cite{Bena:2009qv, ReissnerNord}.

The purpose of this paper is four-fold.  First, we examine the full
supergravity equations of motion using the floating-brane Ansatz, and
show how to obtain the usual BPS and almost-BPS solutions.  In addition, we
find that the linear equations governing these solutions lead to
solutions to the supergravity equations of motion not only when the
base space is hyper-K\"ahler, but also when the base space is merely an
arbitrary four-dimensional Ricci-flat manifold\footnote{Examples of
  such solutions that have Euclidean Schwarzschild and Kerr-Taub-Bolt
  base spaces are explicitly constructed in \cite{Bena:2009qv}.}.  In
hindsight one should have expected this -- after all, having a complex
structure on the base is needed for supersymmetry, but Einstein's
equations only see the Ricci tensor of the base.

Secondly, within the floating-brane Ansatz, we find, after a few
simplifying assumptions, a new class of solutions that solve the
five-dimensional supergravity equations of motion.  The equations
governing these simplified-floating-brane solutions can still be
solved in a linear fashion, but they are more general than both the
BPS and the almost-BPS equations, and reduce  to these in certain
limits. In particular, the simplified-floating-brane solutions have a
four-dimensional base space that does not need to be Ricci-flat.
Rather, to construct these solutions the starting point is a
four-dimensional Euclidean ``electrovac'' solution of the
Einstein-Maxwell equations.  Given such a solution one can turn on
two-form field strengths on various two-cycles, and solve a set of
coupled linear equations to find the remaining two Maxwell field
strengths and two of the warp factors.  One then solves another set of
coupled linear equations to find the rotation vector and the remaining
warp factor. Hence, one can construct full non-BPS solutions of the
five-dimensional supergravity equations of motion starting from any
four-dimensional Euclidean electrovac solution, and solving only {\it
  linear} equations.

The third purpose of this paper is to illustrate the method outlined
above and to construct simplified-floating-brane solutions using
Israel-Wilson geometries as base spaces.  These geometries are a
special class of non-Ricci-flat electrovac solutions that have a
$U(1)$ isometry.  The new equations imply that the functions
determining the magnetic field strengths are no longer harmonic in the
$\IR^3$ base of the Israel-Wilson space, but satisfy a linear system
of coupled differential equations that relate them to some of the warp
factors. We solve this system for the particular example of an
Israel-Wilson base whose fiber degenerates at two locations, and
obtain a solution that, in a certain limit, reduces to a BPS black
hole in Taub-NUT, and in a different limit reduces to a non-BPS black
ring in Taub-NUT. This solution also has a two-cycle with
non-trivial flux, and one can use it to obtain a very large class of
smooth horizonless solutions.

The fourth purpose of this paper is to relate the solutions in our new
class that are constructed using an asymptotically $\IR^3 \times S^1$
Israel-Wilson base space to the known multi-center almost-BPS
solutions in Taub-NUT \cite{Bena:2009ev,recent}.  We find that the two
classes of solutions can be transformed into each other upon applying
the ``spectral flow'' transformation of supergravity solutions with a
$U(1)$ isometry discussed in \cite{Bena:2008wt}. From the perspective
of six-dimensional supergravity (or of the full solution written in a
IIB duality frame where the M2 charges correspond to D1, D5, and P
charges) this transformation mixes the Kaluza-Klein\footnote{That is,
  the $U(1)$ common to the D1 and D5 branes.} $U(1)$ and the $U(1)$ of
the base. For BPS solutions, this spectral flow transformation
re-shuffles the D6, D4, D2 and D0 charges and moduli, but the
resulting solution is still BPS and hence remains in the class of
solutions of \cite{5dsugra,gutowski-reall,Bena:2004de}. However, when
applying this spectral flow transformation to an almost-BPS solution,
the resulting solution is no longer an almost-BPS solution, but is a
simplified-floating-brane solution with an Israel-Wilson base space.

An immediate corollary of this observation is that among within the
floating-brane solutions there exist not only multiple black holes,
but also new smooth horizonless bubbling solutions, that have
non-trivial magnetic fluxes on the two-cycles of the Israel-Wilson
base. Recall that this was not possible for the almost-BPS solutions:
The anti-self-dual flux on the two-cycles of a multi-center Taub-NUT
space is non-normalizable, and does not lead to asymptotically-flat
solutions. Given that solutions in our new class can be obtained by
spectral flow from almost-BPS solutions, it is straightforward to
obtain smooth solutions with non-trivial fluxes by spectrally-flowing
multiple supertubes.

In Section 2 we begin by specifying the floating-brane Ansatz and then
deriving the equations of motion.  For simplicity we work with the
ungauged five-dimensional supergravity action with three $U(1)$ gauge
fields (or the STU model upon reduction to four-dimensions). It would
be interesting to see how difficult it is to extend our solutions to
more general ungauged $U(1)^N$ supergravities.  In Section 3 we first
solve the system of equations as far as possible without making any
additional assumptions beyond the original Ansatz. We then we make
some simple assumptions that lead to a new linear system of equations
that yield new classes of ``simplified-floating-brane'' non-BPS
solutions that have as a base space a four-dimensional
Einstein-Maxwell Euclidean electrovac solution.  These equations,
together with the recipe for constructing simplified-floating-brane
solutions are given in sub-section 3.3.  In Section 4 we solve this
linear system of equations when the base space is an Israel-Wilson
metric, and present an explicit solution describing a non-BPS
D6-D4-D2-D0 black hole in an anti-D6 background. In Section 5 we
discuss the effect of spectral flow on the solutions we build, and
show that the simplified-floating-brane solutions with an
Israel-Wilson base can be mapped to almost-BPS solutions that have a
Gibbons-Hawking base space. In Section 6 we conclude and outline
future directions.

%%%%%%%%%%%%%%%%%%%%%%%%%%%%%%%%%%%%%
\section{Equations of motion}
%%%%%%%%%%%%%%%%%%%%%%%%%%%%%%%%%%%%%

%%%%%%%%%%%%%%%%%%%%%%%%%%%%%
\subsection{Conventions and the floating-brane Ansatz}
%%%%%%%%%%%%%%%%%%%%%%%%%%%%%

We consider $\cN \! = \!  2$, five-dimensional supergravity with three
$U(1)$ gauged fields and we use the conventions of
\cite{Goldstein:2008fq}.  The bosonic action is:
\begin{eqnarray}
  S = \frac {1}{ 2 \kappa_{5}} \int\!\sqrt{-g}\,d^5x \Big( R  -\coeff{1}{2} Q_{IJ} F_{\mu \nu}^I   F^{J \mu \nu} - Q_{IJ} \partial_\mu X^I  \partial^\mu X^J -\coeff {1}{24} C_{IJK} F^I_{ \mu \nu} F^J_{\rho\sigma} A^K_{\lambda} \bar\epsilon^{\mu\nu\rho\sigma\lambda}\Big) \,,
  \label{5daction}
\end{eqnarray}
with $I, J =1,2,3$.  One of the photons lies in the gravity multiplet
and so there are only two vector multiplets and hence only two
independent scalars.  Thus the scalars, $X^I$, satisfy a constraint,
and it is convenient to introduce three other scalar fields, $Z_I$, to
parametrize these two scalars:
\begin{equation}
X^1 X^2 X^3  = 1\,, \qquad  X^1    =\bigg( \frac{Z_2 \, Z_3}{Z_1^2} \bigg)^{1/3} \,, \quad X^2    = \bigg( \frac{Z_1 \, Z_3}{Z_2^2} \bigg)^{1/3} \,,\quad X^3   =\bigg( \frac{Z_1 \, Z_2}{Z_3^2} \bigg)^{1/3}  \,.
\label{XZrelns}
\end{equation}
The scalar kinetic term can be written as:
\begin{equation}
  Q_{IJ} ~=~    \frac{1}{2} \,{\rm diag}\,\big((X^1)^{-2} , (X^2)^{-2},(X^3)^{-2} \big) \,.
\label{scalarkinterm}
\end{equation}
Note that the scalars, $X^I$,   only depend upon the ratios $Z_J/Z_K$ and it is convenient to parametrize the  third independent scalar by:
\begin{equation}
Z ~\equiv~ \big( Z_1 \, Z_2 \, Z_3  \big)^{1/3}   \,.
\label{Zdefn}
\end{equation}

We now use the scalar, $Z$, in the metric Ansatz:
\begin{equation}
ds_5^2 ~=~ -Z^{-2} \,(dt + k)^2 ~+~ Z \, ds_4^2  \,,
\label{metAnsatz}
\end{equation}
where the powers guarantee that $Z$ becomes an independent scalar from the four-dimensional perspective.
We will denote the frames for (\ref{metAnsatz}) by $e^A$, $A=0,1, \dots,4$  and let $\hat e^a$, $a=1, \dots,4$ denote frames for $ds_4^2$.  That is, we take:
\begin{equation}
e^0 ~\equiv~     -Z^{-1} \,(dt + k)\,, \qquad e^a ~\equiv~  Z^{1/2} \,\hat e^a \,.
\label{frames}
\end{equation}

The heart of the ``floating brane'' Ansatz is to relate the metric coefficients and the
scalars to the electrostatic potentials.  The Maxwell Ansatz is thus:
\begin{equation}
A^{(I)}   ~=~  - \varepsilon\, Z_I^{-1}\, (dt +k) + B^{(I)}  \,,
\label{AAnsatz}
\end{equation}
where $B^{(I)}$ is a one-form on the base (with metric $ds_4^2$). The
parameter, $\varepsilon$, will be related to the self-duality or
anti-self-duality of the fields in the solution and is fixed to have
$\varepsilon^2 =1$. Upon uplifting this solutions to
eleven-dimensional supergravity, or M-theory, this Ansatz implies that
M2 brane probes that have the same charge vector as the M2 charge
vector of the solution will have equal and opposite Wess-Zumino and
Born-Infeld terms and hence will feel no force.  Such brane probes may
be placed anywhere in the base and may thus be viewed as ``floating.''

It is convenient to define the field strengths:
\begin{equation}
\Theta^{(I)}    ~\equiv~  d B^{(I)}  ~=~ \coeff{1}{2} \, Z^{-1} \,\Theta^{(I)}_{ab} \, e^a \wedge e^b    ~=~ \coeff{1}{2} \, \Theta^{(I)}_{ab} \, \hat e^a \wedge \hat e^b   \,.
\label{Thetadefn}
\end{equation}
and
\begin{equation}
K   ~\equiv~  d k  ~=~ \coeff{1}{2}  \, (\partial_ \mu \, k_\nu - \partial_ \nu \, k_\mu) \, dx^\mu \wedge dx^\mu ~=~ \coeff{1}{2}  \, K_{ab} \,\hat e^a \wedge \hat e^b   \,.
\label{Kdefn}
\end{equation}
Note that the frame components are defined relative to the frames on $ds_4^2$.

Another consequence of the fact that we have used the same function,
$Z$, in both the metric and the electric potential in (\ref{AAnsatz})
is that the mass of our solutions will always be linear in the
electric (M2) charges, much like the mass of extremal solutions
(although for some orientations the mass may also {\it decrease}
linearly with the charges \cite{Bena:2009qv}).  This also suggests
that our solutions should be essentially extremal, however we have
made no assumptions about the base metric, $ds_4^2$, and the choices
for this will lead to a very large class of non-BPS solutions that
include non-extremal solutions.

 %%%%%%%%%%%%%%%%%%%
\subsection{Einstein's equations}
%%%%%%%%%%%%%%%%%%%

The time ($00$) components of Einstein's equations give:
\begin{equation}
\sum_I \, Z_I^{-1} \hat \nabla^2 Z_I   ~=~ -  \coeff{1}{4}  \, Z^{-3} \, \sum_I \, Z_I \,\Theta^{(I)}_{ab} \, \big( Z_I  \, \Theta^{(I)}_{ab} ~-~ 2\, \varepsilon   \, K_{ab}\big)   \,,
\label{Eins00}
\end{equation}
where $\hat \nabla$ is the covariant derivative in the base metric, $ds_4^2$.

The off-diagonal ($0a$ components) of Einstein's equations give:
\begin{equation}
 \hat \nabla^b K_{ba}   ~=~   \varepsilon \, \sum_I   \, \big(\hat  \nabla^b Z_I\big) \, \Theta^{(I)}_{ba}  \,,
\label{Eins0a}
\end{equation}
or, equivalently,
\begin{equation}
d *_4 K   ~=~   \varepsilon \, \sum_I   \, dZ_I  \, \wedge \, *_4 \Theta^{(I)}  \,.
\label{Eins0aform}
\end{equation}

To give the remaining Einstein's equations it is convenient to define the two-form:
\begin{equation}
\cP   ~ \equiv ~ K   ~-~ \coeff{1}{2} \, \varepsilon \, \sum_{I=1}^3 \, Z_I   \,\Theta^{(I)}  \,.
\label{Pdefn}
\end{equation}
The components  of Einstein's equations on the four-dimensional base are:
\begin{eqnarray}
 \hat R_{ab}   ~-~ \coeff{1}{2}  \hat R\, \delta_{ab}&=&  Z^{-3}\, \bigg[  \cP_{ac} \,  \cP_{bc}   ~-~  \coeff{1}{4}  \, \delta_{ab}  \,   \cP_{cd} \cP_{cd} \nonumber \\
 &&\qquad ~+~
  \coeff{1}{4}  \,  \Big(2\,\sum_I \,  Z_I^2 \,   \Theta^{(I)}_{ac} \,  \Theta^{(I)}_{bc} ~-~     \, \sum_{I,J }\,  Z_I Z_J  \,   \Theta^{(I)}_{ac} \,  \Theta^{(J)}_{bc} \Big)   \nonumber \\
 &&\qquad   ~-~  \coeff{1}{16} \, \delta_{ab}  \, \Big( 2\,\sum_I \,  Z_I^2 \,   \Theta^{(I)}_{cd} \,  \Theta^{(I)}_{cd} ~-~   \sum_{I,J}\,  Z_I Z_J \,   \Theta^{(I)}_{cd} \,  \Theta^{(J)}_{cd} \Big)\, \bigg]   \,,
\label{Einsab}
\end{eqnarray}
where $\hat R_{ab}$ and $\hat R$ are the Ricci tensor and Ricci scalar of the base metric, $ds_4^2$.  Note that these equations imply that the Ricci scalar of the base must  vanish:
\be
\hat R =0 \,.
\ee

%%%%%%%%%%%%%%%%%%%
\subsection{Scalar equations}
%%%%%%%%%%%%%%%%%%%

The scalar equations of motion yield equations for the ratios of the $Z_I$.  For example:
\begin{eqnarray}
 Z_1^{-1} \hat \nabla^2 Z_1 ~-~  Z_3^{-1} \hat \nabla^2 Z_3    &=&   \coeff{1}{2}  \, Z^{-3} \, \bigg[ Z_1 \,\Theta^{(1)}_{ab} \, \big( Z_1  \, \Theta^{(1)}_{ab} ~-~ 2\, \varepsilon   \, K_{ab}\big) \\ && \qquad\qquad ~-~ Z_3 \,\Theta^{(3)}_{ab} \, \big( Z_3  \, \Theta^{(3)}_{ab} ~-~ 2\, \varepsilon   \, K_{ab}\big)  \bigg]    \,.
\label{scalar13}
\end{eqnarray}

When combined with (\ref{Eins00}) one gets:
\begin{eqnarray}
 \hat \nabla^2 Z_I   &=&   -\coeff{1}{4}  \, Z_J^{-1} Z_K^{-1}  \bigg[ Z_J \,\Theta^{(J)}_{ab} \, \big( Z_J  \, \Theta^{(J)}_{ab} ~-~ 2\, \varepsilon   \, K_{ab}\big) \\ &&  ~+~ Z_K \,\Theta^{(K)}_{ab} \, \big( Z_K  \, \Theta^{(K)}_{ab} ~-~ 2\, \varepsilon   \, K_{ab}\big)  ~-~ Z_I \,\Theta^{(I)}_{ab} \, \big( Z_I  \, \Theta^{(I)}_{ab} ~-~ 2\, \varepsilon   \, K_{ab}\big) \bigg]    \,,
\label{ZIeqn}
\end{eqnarray}
where $\{I,J,K\} =\{1,2,3\}$ are all distinct and these indices are not summed.

 %%%%%%%%%%%%%%%%%%%
\subsection{Maxwell equations}
%%%%%%%%%%%%%%%%%%%

To give the Maxwell equations it is convenient to define:
\begin{equation}
{\cR}^{(I)}_\pm  ~\equiv~     \coeff{1}{2} \, \varepsilon  \,Z_I\,\big(\Theta^{(I)} ~\pm~ \varepsilon  *_4 \Theta^{(I)} \big)  \,, \qquad
\cP_\pm ~ \equiv ~ \coeff{1}{2} \, \big(K ~\pm~ \varepsilon  *_4 K \big) ~-~  \coeff{1}{2} \, \sum_{M=1}^3 \, {\cR}^{(M)}_\pm
\label{PRdefn}
\end{equation}
with no sum on $I$.  Note that $\cP = \cP_+ + \cP_-$. The parameter, $\varepsilon$, satisfies $\varepsilon^2 =1$ and so determines whether these combinations are self-dual or anti-self-dual.

The Maxwell equations are:
\begin{equation}
d *_5 \big(Q_{IJ} \,F^{J}\big)  ~=~     \coeff{1}{4} \, C_{IJK}\,   F^J \wedge F^K \,,
\label{Max1}
\end{equation}
with $F^I = d A^I$.  Using the Ansatz (\ref{AAnsatz}) one obtains two types of terms:  (i) a four form on the four-dimensional base and (ii) $e^0$ wedged into a three form on the four-dimensional base.  The former generates the following equations for  $\hat \nabla^2 Z_I $:
\begin{equation}
 \hat \nabla^2 Z_I ~=~  \varepsilon  *_4 \bigg[ \Theta^{(J)} \wedge \Theta^{(K)}    ~+~Z^{-3} Z_I \,K \wedge \, \Big( K +  \varepsilon *_4 K + 2{\cR}^{(I)}_- -  \varepsilon  \, \sum_{M=1}^3 Z_M   \Theta^{(M)}   \Big) \bigg]    \,.
\label{ZIeqnother}
\end{equation}
Combining this with (\ref{ZIeqn}) one obtains three algebraic constraints on the forms  $\cP_+$ and ${\cR}^{(M)}_\pm$:
\begin{equation}
\cP_+ \wedge \cP_+  ~+~  \cP_+ \wedge  {\cR}^{(I)}_+ ~+~\coeff{1}{4}\, \big( {\cR}^{(I)}_- - {\cR}^{(J)}_- + {\cR}^{(K)}_- \big) \wedge  \big( {\cR}^{(I)}_- + {\cR}^{(J)}_- - {\cR}^{(K)}_- \big)  ~=~     0 \,,
\label{Max2}
\end{equation}
where, once again,  $\{I,J,K\} =\{1,2,3\}$ are all distinct and these indices are not summed.

The second set of Maxwell equations can be written as:
\begin{eqnarray}
d\Big(Z^{-3} Z_I \, \Big( K +  \varepsilon *_4 K + 2{\cR}^{(I)}_- -  \varepsilon  \, \sum_{M=1}^3 Z_M   \Theta^{(M)}   \Big)\Big) \,=\, 0  \,,
\label{Max3}
\end{eqnarray}
where the index $I$ isn't summed.

%%%%%%%%%%%%%%%%%%%%%%%%%%%%%%%%%%%%%
\section{Solving the system}
%%%%%%%%%%%%%%%%%%%%%%%%%%%%%%%%%%%%%

%%%%%%%%%%%%%%%%%%%
\subsection{General results}
%%%%%%%%%%%%%%%%%%%

Using the equations of motion one can easily show that:
\begin{equation}
d  \big((Z_1 Z_2 Z_3)^{-1} \cP_+  \big)~=~   0 \,,
\label{GenCond1}
\end{equation}
and hence one may write
\begin{equation}
\cP_+ ~=~ (Z_1 Z_2 Z_3)\, \omega_+^{(0)}\,,
\label{PCond1}
\end{equation}
where $\omega_+^{(0)}$ is harmonic.

One can simplify some of the Maxwell equations by introducing some additional forms, $\omega^{(I)}_-$ defined by:
\begin{equation}
\coeff{1}{2} \, \varepsilon\, \big(\Theta^{(I)} - \varepsilon  *_4 \Theta^{(I)} \big)  ~\equiv~ C_{IJK} \, Z_J \, \omega^{(K)}_- \,.
\label{omdefns}
\end{equation}
Since $Z_1 Z_2 Z_3 \ne 0$, this transformation is invertible and so we have made no additional assumptions.
In terms of these new $\varepsilon$-anti-self-dual forms, the Maxwell equations (\ref{Max3}) become:
\begin{equation}
d *_4  \omega^{(I)}_-  ~=~ (Z_1 Z_2 Z_3)^{-1} dZ_I \wedge\, \cP_+ \,.
\label{Max4}
\end{equation}
and  (\ref{Max2}) becomes:
\begin{equation}
\cP_+ \wedge \cP_+  ~+~  \cP_+ \wedge  {\cR}^{(I)}_+ ~+~ (Z_1 Z_2 Z_3)  \,Z_I\,  \omega^{(J)}_-   \wedge\omega^{(K)}_- ~=~     0 \,,
\label{Max5}
\end{equation}
where $\{I,J,K\} =\{1,2,3\}$ are all distinct and these indices are not summed.

To simplify Einstein's equations, introduce the function, $\cT_{ab}$,  of a pair of two forms that is defined by:
\begin{equation}
\cT_{ab} (X,Y)  ~\equiv~   \coeff{1}{2} \,\big( X_{ac} \,  Y_{bc} ~+~ X_{bc} \,  Y_{ac}   \big) ~-~  \coeff{1}{4} \, \delta_{ab}  \,    X_{cd} \,  Y_{cd}  \,.
\label{Tdefn}
\end{equation}
In particular, $\cT_{ab}(F,F)$ is the energy momentum tensor associated with the Maxwell field, $F$.  Note that if $X_\pm$ and $Y_\pm$ are the self-dual and anti-self dual parts of $X$ and $Y$, then
\begin{equation}
\cT_{ab} (X_\pm,Y_\pm)  ~=~   0\,, \qquad  \cT_{ab} (X ,Y )  ~=~   \cT_{ab} (X_+,Y_-)  ~+~    \cT_{ab} (X_-,Y_+) \,.
\label{Trelns}
\end{equation}
Using this in the Einstein equations (\ref{Einsab}), one obtains:
\begin{equation}
 \hat R_{ab}  ~=~ 2\, Z^{-3}\, \cT_{ab}(\cP_+ ,\cP_-) ~-~\sum_{I=1}^3\, \cT_{ab} \big(\coeff{1}{2} \, \varepsilon\, \big(\Theta^{(I)} + \varepsilon  *_4 \Theta^{(I)} \big)   ,  \omega^{(I)}_-\big)   \,.
\label{Einsred}
\end{equation}
Thus far we have made no assumptions other than our floating brane Ansatz.

%%%%%%%%%%%%%%%%%%%
\subsection{A simple assumption}
%%%%%%%%%%%%%%%%%%%

The equations of motion dramatically simplify if one takes:
\begin{equation}
\cP_+   ~\equiv~   0 \,,
\label{KillPplus}
\end{equation}
which is, of course, consistent with (\ref{GenCond1}) and thus with the equations of motion. We will henceforth assume that (\ref{KillPplus}) is true.

One then finds from (\ref{Max4}) and  (\ref{Max5}) that the  forms $\omega^{(I)}_- $ must be harmonic and satisfy
\begin{equation}
\omega^{(I)}_-   \wedge \omega^{(J)}_- ~=~     0 \,, \quad {I \ne J} \,.
\label{vanwedge}
\end{equation}
There are two obvious ways to satisfy this condition: \\
$\bullet$ (i) Take $\omega^{(1)}_-  =\omega^{(2)}_- =0$ and $\omega^{(3)}_- $ to be an arbitrary $\varepsilon$-anti-self-dual harmonic form. \\
$\bullet$ (ii)  Take the manifold to be hyper-K\"ahler and let each of the  $\omega^{(I)}_- $ be a constant multiple  of one the three harmonic two forms associated with the three complex structures\footnote{There might be an interesting generalization of (ii)  to quaternionic-K\"ahler spaces.}.

Continuing with the implications of (\ref{KillPplus}), one finds that the equations for the scalars (\ref{ZIeqnother}) reduce to:
\begin{equation}
 \hat \nabla^2 Z_I ~=~ \varepsilon \, *_4  \Big[ \Theta^{(J)} \wedge \Theta^{(K)}   ~-~  \varepsilon  \, \omega^{(I)}_- \wedge( K -  \varepsilon *_4 K)  \Big]    \,,
\label{ZIeqnsimp}
\end{equation}
and Einstein's equations collapse to
\begin{equation}
 \hat R_{ab}  ~=~  -\coeff{1}{2} \, \varepsilon\, \sum_{I=1}^3\, \cT_{ab} \big(\big(\Theta^{(I)} + \varepsilon  *_4 \Theta^{(I)} \big)   ,  \omega^{(I)}_-\big)   \,.
\label{Einsimp}
\end{equation}
Note that the Ricci tensor depends only upon the four-dimensional electromagnetic fluxes.

%%%%%%%%%%%%%%%%%%%
\subsection{A further assumption and a linear system}
\label{linsys}
%%%%%%%%%%%%%%%%%%%

We now make a further assumption, that condition (i) above is satisfied: Hence  $\omega^{(1)}_-  =\omega^{(2)}_- =0$ and $\omega^{(3)}_- $ is an arbitrary $\varepsilon$-anti-self-dual harmonic form.  Then the equations become
\begin{eqnarray}
\big(\Theta^{(1)} - \varepsilon  *_4 \Theta^{(1)} \big)  &=&  2 \, \varepsilon\, Z_2 \, \omega^{(3)}_- \,, \qquad
\big(\Theta^{(2)} - \varepsilon  *_4 \Theta^{(2)} \big)  ~=~  2 \, \varepsilon\, Z_1 \, \omega^{(3)}_- \,, \nonumber \\
\big(\Theta^{(3)} - \varepsilon  *_4 \Theta^{(3)} \big)  &=&  0 \,.
\label{Vsimpforms}
\end{eqnarray}
Thus $\Theta^{(3)}$ is a harmonic, $\varepsilon $-self-dual two form.

The background geometry must be chosen so that
\begin{equation}
 \hat R_{ab}  ~=~  - \varepsilon\,  \cT_{ab} \big( \Theta^{(3)}   \,,  \omega^{(3)}_-\big)   ~\equiv~   \coeff{1}{2} \,\big( \cF_{ac} \,  \cF_{bc}   ~-~  \coeff{1}{4}  \, \delta_{ab}  \,   \cF_{cd} \cF_{cd} \big) \,,
\label{EinVsimp}
\end{equation}
 where $\cF$ is defined by
\begin{equation}
\cF   ~\equiv~  \Theta^{(3)}   ~-~   \varepsilon\, \omega^{(3)}_- \,.
\label{cFdefn}
\end{equation}
Note that this Maxwell field must be harmonic.

To find a full solution of the supergravity equations of motion one
must start from a Euclidean ``electrovac'' solution to $U(1)$
Einstein-Maxwell theory. The metric of this solution will be the base
metric of the full geometry, and the self- and anti-self-dual parts of
the electrovac Maxwell field determine $ \Theta^{(3)}$ and
$\omega^{(3)}_-$. Note that both these forms must be closed, as a
consequence of the Maxwell equations and Bianchi identities for $\cF$.
They will therefore automatically satisfy equations (\ref{Vsimpforms}) and
(\ref{Max4}) under assumption (i). Conversely, given any solution to
our equations, one can always repackage $ \Theta^{(3)}$ and
$\omega^{(3)}_-$ into a Maxwell field that satisfies (\ref{EinVsimp}),
and obtain an electrovac solution.

Given  $ \Theta^{(3)}$ and $\omega^{(3)}_-$, we then need to solve the following pairs of equations:
\begin{eqnarray}
 \hat \nabla^2 Z_1 &=&  \varepsilon \, *_4  \big[ \Theta^{(2)} \wedge \Theta^{(3)}   \big]    \,,  \qquad \big(\Theta^{(2)} - \varepsilon  *_4 \Theta^{(2)} \big)  ~=~  2 \, \varepsilon\, Z_1 \, \omega^{(3)}_- \,;  \label{Zsimpform1} \\
\hat \nabla^2 Z_2&=& \varepsilon \, *_4  \big[ \Theta^{(1)} \wedge \Theta^{(3)}   \big]  \,,  \qquad
 \big(\Theta^{(1)} - \varepsilon  *_4 \Theta^{(1)} \big)    ~=~ 2 \, \varepsilon\, Z_2 \, \omega^{(3)}_- \,.
 \label{Zsimpform2}
\end{eqnarray}
Since $ \Theta^{(3)}$ and $\omega^{(3)}_-$ are already known,
(\ref{Zsimpform1}) represents a coupled {\it linear} system for $
\Theta^{(2)}$ and $Z_1$ and (\ref{Zsimpform2}) represents a coupled
{\it linear} system for $ \Theta^{(1)}$ and $Z_2$.  In solving these
systems one should, of course, remember that the $ \Theta^{(I)}$
should also satisfy the (linear) Bianchi identities $d\Theta^{(1)}
=0$.

Once one knows the solutions of the equations above, one must solve the equations for $Z_3$ and $K= dk$,  which, amazingly enough, are also linear:
\begin{eqnarray}
 K   ~+~   \varepsilon *_4 K    &=&  \coeff{1}{2} \,\varepsilon \sum_I \, Z_I \,  \big(\Theta^{(I)} + \varepsilon    *_4 \Theta^{(I)} \big)    \,, \qquad  \label{ksimpforms}  \\
  \hat \nabla^2 Z_3  &=&  \varepsilon \, *_4  \big[ \Theta^{(1)} \wedge \Theta^{(2)}   ~-~  \varepsilon  \, \omega^{(3)}_- \wedge( K -  \varepsilon *_4 K)  \big]    \,.
\label{Zsimpforms}
\end{eqnarray}
Hence, starting from an Euclidean electrovac solution one can build a
full solution of five-dimensional $U(1)^3$ ungauged supergravity by
following a linear procedure, much like one does for BPS and
almost-BPS solutions. Note however that our class of solutions is much
larger, and includes the BPS and almost-BPS solutions.  The
latter merge by restricting the electric fields of the Euclidean
electrovac solution to be self- or anti-self-dual, and thus the base space
becomes Ricci-flat. We now explain in detail how this happens.

If $\omega_-^{(3)}=0$,  then one has
\begin{equation}
\Theta^{(I)} ~=~  \varepsilon  *_4 \Theta^{(I)} \,,  \qquad \hat \nabla^2 Z_I ~=~ \varepsilon \, \coeff{1}{2}\, C_{IJK}   *_4 \Theta^{(J)} \wedge \Theta^{(K)} \,,  \qquad \cP_+   ~=~   0  \,.
\label{BPSeqns}
\end{equation}
For $ \varepsilon=1$, these are just the BPS equations of
\cite{Bena:2004de,gutowski-reall}. For $ \varepsilon=-1$ they become
the almost-BPS equations of \cite{Goldstein:2008fq}. Nevertheless,
note that (\ref{Einsimp}) implies that the base space of these
solutions does not need to be hyper-K\"ahler, but only Ricci flat:
\begin{equation}
 \hat R_{ab}  ~=~ 0   \,.
\label{Ricciflat}
\end{equation}
Hence, both the BPS and the almost-BPS equations allow for non-trivial
non-BPS solutions with non-hyper-K\"ahler bases. Examples of such
solutions are constructed in \cite{Bena:2009qv}. The fact that the
base only needs to be Ricci-flat is not so strange: The hyper-K\"ahler
condition originally arose because one wanted to preserve
supersymmetry, however, Einstein's equations, (\ref{Einsred}), only
care about the Ricci tensor of the base. It is also interesting to
note that there also exist solutions with Ricci-flat bases that have
$\omega_-^{(3)}\neq0$ and $\Theta^{(3)}=0$. These solutions can in
principle be obtained by recycling the known BPS and almost-BPS
solutions, and their physics is worth exploring further.

%%%%%%%%%%%%%%%%%%%%%%%%%%%%%%%%%%%%%
\section{Israel-Wilson metrics} \label{IWmetrics}
%%%%%%%%%%%%%%%%%%%%%%%%%%%%%%%%%%%%%

We now examine, in more detail, the linear system of equations found
in Section \ref{linsys}, and solve it for a special class of
electrovac base spaces that have a translational $U(1)$ isometry: the
Israel-Wilson spaces. From a four-dimensional perspective, the
highly non-trivial particular solution we find describes a non-BPS
two-centered solution where one of the centers is a locally-BPS
D6-D4-D2-D0 black hole and the other center is a
$\overline{\mathrm{D}6}$ brane.

%%%%%%%%%%%%%%%%%%%%%%
\subsection{The Israel-Wilson background}
%%%%%%%%%%%%%%%%%%%%%%

The starting ingredient for constructing non-trivial solutions using
the procedure outlined above is an Euclidean electrovac solution that
satisfies (\ref{EinVsimp}) and that has a non-trivial harmonic form,
$\omega_-^{(3)}$.  An interesting choice for such a background is an
Israel-Wilson (IW) metric \cite{Israel:1972vx,
  Whitt:1984wk,Yuille:1987vw,Dunajski:2006vs}:
\begin{equation}
ds^2_4 = (V_+ \,  V_-)^{-1}(d\psi+ \vec A\cdot d \vec y )^2 + (V_+ \,  V_-) (dy_1^2 +dy_2^2 + dy_3^2) \,,
\end{equation}
where
\begin{equation}
\vec \nabla \times \vec A = V_- \vec \nabla V_+  -V_+ \vec \nabla V_-\,,
\label{Adefn}
\end{equation}
and the functions $V_\pm$  are required to be harmonic on the $\IR^3$ base.
Introducing the frames:
\begin{equation}
\hat e^1~=~  (V_+ \,  V_-)^{-{1\over 2}}\, (d\psi ~+~  \vec A\cdot d \vec y) \,,
\qquad \hat e^{a+1} ~=~  (V_+ \,  V_-)^{1\over 2}\, dy^a \,, \quad a=1,2,3 \,,
\label{IWframes}
\end{equation}
the background Maxwell field is given by \cite{Dunajski:2006vs}:
\begin{eqnarray}
\cF &\equiv&\coeff{1}{2} \,\cF_{ab} \hat  e^a \wedge  \hat  e^b  \nonumber \\
&=&   \big[ \partial_a \big(V_+^{-1} - V_-^{-1} \big) \big] \, e^1 \wedge \hat  e^{a+1} ~+~
\coeff{1}{2} \,\epsilon_{abc}  \big[ \partial_a \big(V_+^{-1} + V_-^{-1} \big) \big]  e^{b+1} \wedge \hat  e^{c+1}    \,.
\label{IWFform}
\end{eqnarray}
This background then satisfies equation (\ref{EinVsimp}).

%%%%%%%%%%%%%%%%%%%%%%
\subsection{Harmonic forms}
%%%%%%%%%%%%%%%%%%%%%%

Define the sets of two-forms:
\begin{equation}
\Omega_\pm^{(a)} ~\equiv~ \hat e^1  \wedge \hat
e^{a+1} ~\pm~ \coeff{1}{2}\, \epsilon_{abc}\,\hat e^{b+1}  \wedge
\hat e^{c+1} \,, \qquad a =1,2,3\,.
\label{twoforms}
\end{equation}
The Maxwell field of the Israel-Wilson solution is then:
\begin{equation}
\cF ~=~
 \big(\partial_a \big(V_+^{-1} \big) \big)  \,  \Omega_+^{(a)}  ~-~ \big(\partial_a \big(V_-^{-1} \big) \big)  \,  \Omega_-^{(a)}   \,,
\label{IWFformsimp}
\end{equation}
from which one can read off (up to an irrelevant, overall sign) the harmonic forms, $\Theta^{(3)}$ and $\omega^{(3)}_-$, using (\ref{cFdefn}). However, it is easy to see that (\ref{IWFform}) is not the most general Maxwell field one can have for this base. Introducing two arbitrary harmonic functions $K_\pm$ on $\mathbb{R}^3$, the two-forms:
\begin{equation}
\Theta_\pm  ~ \equiv~ - \sum_{a=1}^3 \, \big(\partial_a \big( V_\pm^{-1}\, K_ \pm \big)\big) \,
\Omega_\pm^{(a)}
\label{harmtwoform}
\end{equation}
are also harmonic and self-dual, or anti-self-dual respectively.  These forms have (local) potentials:
\begin{equation}
a_\pm  ~=~  {K_\pm \over V_\pm} (d\psi+   \vec A\cdot d \vec y) ~+~  \vec b_\pm  \cdot d \vec y \,,\qquad  \vec \nabla \times \vec b_\pm =  \pm \left( K_\pm  \vec \nabla V_\mp  - V_\mp \vec \nabla K_\pm  \right) \,.
\end{equation}

From now on we choose $\varepsilon=1$. The equations for ($\varepsilon=-1$) can be  simply obtained by exchanging $V_+$ and $V_-$. The two-form, $\Theta^{(3)}$, is then self-dual while $\omega_-^{(3)}$ is  anti-self-dual.

One can try to obtain a more general solution for $\Theta^{(3)}$ and $\omega^{(3)}_-$ by taking:
\begin{equation}
 \Theta^{(3)} ~=~  d \Big( {K_+\over V_+} (d\psi+ \vec A) + \vec b_+  \Big) \,,  \qquad  \omega_-^{(3)} ~=~  d \Big( {K_-\over V_-} (d\psi+ \vec A) + \vec b_- \Big) \,.
\end{equation}
The Einstein-Maxwell electrovac equations (\ref{EinVsimp}) are then solved if, and only if
\begin{equation} \label{eincond}
\partial_i \Big({ K_+ \over V_+}\Big) \partial_j \Big({K_- \over V_-}\Big) =  \big(\partial_i V_+^{-1}\big) \,\big(\partial_j V_-^{-1}\big)
\end{equation}
for $i,j=1,2,3$. Hence, one can apparently obtain a more general electrovac Israel-Wilson base by using the solutions to this equation:
\begin{equation}
K_- ~=~  \beta \,V_- - \alpha  \,,\quad K_+ ~=~  \delta \,V_+ - \gamma \,,
 \label{relaKV}
\end{equation}
with $\alpha,\beta, \gamma, \delta$ constants satisfying the constraint $\alpha\gamma=1$. However, one can easily see that $\beta $ and  $\delta$ are ``pure gauge'' constants, since they make no contribution to the Maxwell fields (\ref{harmtwoform}). We therefore set $\beta =0$, which implies that $K_-$ is constant. We could, of course, do the same with $K_+$, however, we will find it useful in the   next sub-section to keep $\delta\neq0$.

We should also note that the foregoing discussion no longer applies if either $V_-$ or $V_+$ are constant, because the solutions to (\ref{eincond}) are then different from those in (\ref{relaKV}). We will partially address this situation later in the paper, and we leave a more general analysis for further investigation. Given that $K_-=-\alpha$, the two-form $\omega_-^{(3)}$ is a constant multiple of the natural anti-self-dual two-form on the Israel-Wilson base space, ${\displaystyle \big(\partial_a \big(V_-^{-1} \big) \big)  \,  \Omega_-^{(a)} }$.

%%%%%%%%%%%%%%%%%%%%%%%%%%%
\subsection{The linear system}
%%%%%%%%%%%%%%%%%%%%%%%%%%%

We now solve the linear system for the other fields. We write $\Theta^{(1)}$ and $\Theta^{(2)}$ in the form:
\begin{equation}
 \Theta^{(1)}~=~  d\left( {K_1\over V_+} (d\psi+  A) +  b_1 \right) \,, \qquad  \Theta^{(2)} ~=~  d\left( {K_2\over V_+} (d\psi+  A) + b_2  \right)
\end{equation}
where $K_1$, $K_2$, $b_1$ and $b_2$ are unknown functions and one-forms on the $\IR^3$ base and determine the dipole charges of the solution. Writing the equations (\ref{Zsimpform1}) in the IW base, we obtain
\begin{eqnarray} \label{sysK2Z11}
 \nabla^2 K_2 &=& {2 \alpha \over V_-} \, \vec \nabla \cdot \left({V_+ \over V_-}Z_1 \vec \nabla V_-\right)\,, \\ \label{sysK2Z12}
 \nabla^2  Z_1 &=&  V_- \nabla^2  \Big({K_2K_+ \over V_+} \Big) ~-~  2 \alpha \, \vec \nabla \cdot \Big({Z_1 K_+ \over V_-}  \vec \nabla V_-\Big)\,,
\end{eqnarray}
with $b_2$ given by
\begin{equation}
\vec \nabla \times \vec b_2 ~=~ - V_- \vec \nabla  K_2 + K_2 \vec \nabla  V_-  + 2 \alpha \, {V_+  \over V_-} Z_1 \vec \nabla V_- \,.
\end{equation}
The corresponding system for $Z_2$ and $\Theta^{(1)}$ is:
\begin{eqnarray} \label{sysK1Z21}
 \nabla^2 K_1 &=& {2 \alpha \over V_-} \, \vec \nabla \cdot \left({V_+ \over V_-}Z_2 \vec \nabla V_-\right)\,, \\ \label{sysK1Z22}
 \nabla^2  Z_2 &=&  V_- \nabla^2  \Big({K_1K_+ \over V_+} \Big) ~-~  2 \alpha \, \vec \nabla \cdot \Big({Z_2 K_+ \over V_-}  \vec \nabla V_-\Big)\,, \\
\vec \nabla \times \vec b_1 &=& - V_- \vec \nabla  K_1 + K_1 \vec \nabla  V_-  + 2 \alpha \, {V_+  \over V_-} Z_2 \vec \nabla V_- \,.
\end{eqnarray}

\medskip

We also need the equation for the last warp factor $Z_3$ and the angular momentum $k$. We decompose $k$ as usual:
\be
k = \mu (d\psi +  A) +  \omega \,.
\ee
Equations (\ref{ksimpforms}) and (\ref{Zsimpforms}) for $\mu$ and $Z_3$ then give
\bea \label{sysK3mu1}
 \nabla^2  Z_3 &=&  V_- \nabla^2  \Big({K_1K_2 \over V_+} \Big) ~-~  2 \alpha \, \vec \nabla \cdot \Big({Z_1K_1 + Z_2 K_2 \over V_-}  \vec \nabla V_-\Big) \\ \nonumber
 &&+ ~ 4 \alpha {V_+ \over V_-} \, \vec \nabla \cdot \Big( \mu \vec \nabla V_-\Big) - 2 \alpha {V_+ Z_I \over V_-} \, \vec \nabla \cdot \Big( {K_I \over V_+ }\vec \nabla V_-\Big) \\ \nonumber
 &&+~  2 \alpha^2 V_+ Z_1 Z_2 \, \nabla^2\Big({1\over V_-}\Big) \,, \\ \label{sysK3mu2}
 \nabla^2(V_- \mu) &=& {1 \over V_+} \vec \nabla \cdot \Big( V_- V_+ Z_I \vec \nabla \Big({K_I \over V_+}\Big)\Big) - {2 \alpha \over V_+} \vec \nabla \cdot \Big( {V_+ Z_1 Z_2 \over V_-}  \vec \nabla V_-\Big)
 \label{muVeqn}
\eea
where $\omega$ is given by:
\be
 \vec \nabla \times \vec \omega = V_+^2 \vec \nabla \big({V_-\over V_+}\mu \big) - V_+ V_- Z_I \vec \nabla \big({K_I\over V_+}\big) + 2 \alpha {V_+ Z_1 Z_2 \over V_-} \vec \nabla V_- \,.
\label{omeqn}
\ee
As usual, (\ref{muVeqn}) is the integrability equation for (\ref{omeqn}).

%%%%%%%%%%%%%%%%%%%%%%
\subsection{An explicit example: a non-BPS black hole in an Israel-Wilson metric}
%%%%%%%%%%%%%%%%%%%%%%

We now have all the tools to find explicit solutions with an
Israel-Wilson base space. Here we will present an M-theory solution
that corresponds, in type IIA string theory, to a $ {\mathrm{D}6}
{\mathrm{D}4}^3 {\mathrm{D}2}^3 {\mathrm{D}0}$ black hole in a
$\overline{\mathrm{D}6}$ background. We parameterize the flat,
three-dimensional $\mathbb{R}^3$ base space using spherical
coordinates $(r,\theta,\phi)$ and put the $\overline{\mathrm{D}6}$
brane at the origin of the space and the black hole at a distance $R$
from the origin. We denote polar coordinates centered at the black
hole position by $(\Sigma,\theta_\Sigma)$. Their relation to the polar
coordinates $(r,\theta)$ centered at the origin is:
\be
\Sigma = \sqrt{r^2 + R^2 - 2 r R \cos\theta}\,,\qquad \cos\theta_\Sigma = {r\cos\theta-R\over \Sigma}\,.
\label{polarSigma}
\ee
For $V_+=1$, we want the space to be Taub-NUT, and thus we take $V_-$ to be
\be
 V_- = 1 + {Q_{\bar{6}} \over r} \,.
\ee
The parameter, $Q_{\bar{6}}$, is $\overline{\mathrm{D6}}$ or the
$\overline{\mathrm{KK}}$-monopole charge of the space\footnote{We will
  explain below why we refer to this as $\overline{\mathrm{D6}}$ and
  not as D6 charge.}. The function, $K_+$, is harmonic and corresponds
to one of the M5 charges of the solution:
\be
 K_+ = K_3 = { d_3 \over \Sigma } \,.
\ee
For convenience, we will change notation throughout the rest of the
paper, and refer to $K_-$ as $K_3$. The relation (\ref{relaKV}) then
forces $V_+$ to have a pole at the black hole location. Assuming the
space to be asymptotically flat (asymptotic to $\IR^3\times S^1$)
means that the constant in $V_+$ to be finite, and we set it to 1 for
convenience. Hence,
\be \label{relaKV2}
 V_+ = 1 + \alpha K_+ = 1 + {\alpha \, d_3 \over \Sigma} \equiv 1 + {Q_6 \over \Sigma}\,,
\ee
where $\alpha$ was introduced in (\ref{relaKV}), and we have defined
$Q_6\equiv\alpha d_3$. Thus, the black hole has a finite D6 (or KKm)
charge. The associated vector fields are
\bea
 A &=&  Q_6 {r \cos \theta - R \over \Sigma} d\phi + Q_6 \, Q_{\bar{6}} \, {r  - R \cos \theta \over \Sigma} d\phi - Q_{\bar{6}} \cos \theta d\phi\,, \\
 b_3 &=& - d_3 {r \cos \theta - R \over \Sigma} d\phi - Q_{\bar{6}} \, d_3 \, {r  - R \cos \theta \over \Sigma} d\phi
\eea

The system (\ref{sysK2Z11}) and (\ref{sysK2Z12}) is not completely
straightforward to solve, but, as explained in the previous section,
it is linear in the unknowns $K_2$ and $Z_1$. We find the following
solution:
\bea
 Z_1 &=& {1 \over V_+} \left( 1 + {Q_1\over \Sigma}+ {d_2 d_3 \over \Sigma^2} \Bigl(1+ {Q_{\bar{6}} r\over R^2}\Bigr) \right) \,, \\
 K_2 &=& V_+ \left( {d_2 \over \Sigma} - \alpha \, { Z_1 \over V_- } \right) \,,
\eea
and similarly for $Z_2$ and $K_1$. Here we have also introduced the
dipole charge $d_2$ associated to $K_2$, and the electric charge $Q_1$
of the hole, associated to $Z_1$. The vector field $b_2$ is then given
by
\be
 b_2 = -(d_2 - \alpha Q_1) {r \cos \theta - R \over \Sigma} d\phi - Q_{\bar{6}} \, d_2 \, {r  - R \cos \theta \over \Sigma} d\phi + d_2 Q_6 Q_{\bar{6}} {\cos \theta \over \Sigma^2} d\phi \,.
\ee

The solution to the last system of equations, (\ref{sysK3mu1}) and
(\ref{sysK3mu2}), is:
\bea
 \mu &=& {1 \over V_+ V_-} \Big(( {m\over \Sigma}+ {\tilde m\over r }+{V_- (d_1+d_2+d_3) \over 2 \Sigma}+{ Q_I d_I\over 2 \Sigma^2}+Q_{\bar{6}} Q_I d_I
{\cos\theta\over 2 R \Sigma^2}\nn
&&+ {C_{IJK}\over 6}\,d_I d_J d_K
\Bigl[\Bigl(1+{Q_{\bar{6}}^2\over R^2} \Bigr) \Bigl( {r\cos\theta\over
  R\, \Sigma^3}+\lambda {r\cos\theta-R\over R\, \Sigma^3}\Bigr)+
Q_{\bar{6}} {3 r^2 + R^2\over 2 R^2 r \,\Sigma^3}\Bigr] \Big)  \nn && -\alpha {Z_1 Z_2 \over V_-} \,, \\
 Z_3 &=& V_+ \left( 1 + {Q_3\over \Sigma}+ {d_1 d_2 \over \Sigma^2} \Bigl(1+ {Q_{\bar{6}} r\over R^2}\Bigr) \right)- 2 \alpha V_+ \mu - \alpha^2 { V_+ Z_1 Z_2 \over V_-} \,,
\eea
and
\bea
 \omega &=& -\Bigl[\kappa- m {r \cos\theta- R\over \Sigma} -{\tilde
  m} \cos\theta + {d_1+d_2+d_3\over 2}{r\cos\theta-R\over \Sigma}+Q_{\bar{6}}{d_1+d_2+d_3\over 2}{r-R\cos\theta\over R \Sigma} \nn
&& + Q_{\bar{6}} Q_I d_I
{r\sin^2\theta\over 2 R \,\Sigma^2} + \Bigl(1+{Q_{\bar{6}}^2\over R^2}
\Bigr) {C_{IJK}\over 6}\,d_I d_J d_K (1+\lambda) {r^2 \sin^2\theta\over
  R\, \Sigma^3} \nn
&&+ Q_{\bar{6}} {C_{IJK}\over 6}\,d_I d_J d_K {r (3 R^2 +
  r^2)- R (3 r^2 + R^2) \cos\theta\over 2 R^3\, \Sigma^3}\Bigr]
d\phi\,.
\eea
The constants $m$, ${\tilde m}$, $\kappa$ and $\lambda$ represent homogeneous solutions that are fixed by regularity:
\bea
m &=& \Bigl(1+{Q_{\bar{6}}\over R}\Bigr){d_1+d_2+d_3\over 2}+ {C_{IJK}\over 6}{Q_{\bar{6}} \,d_I d_J d_K\over 2 R^3}\,, \nn
{\tilde m}&=& \kappa = -Q_{\bar{6}} \Bigl({d_1+d_2+d_3\over 2 R}+ {C_{IJK}\over 6}{d_I d_J d_K\over 2 R^3}\Bigr)\,.
\label{regularityring} \\ \nonumber
\lambda &=& -{R^2\over R^2 + Q_{\bar{6}}^2}\,.
\label{alphavalue}
\eea
The reason for this regularity conditions will become clear in the next section.

%%%%%%%%%%%%%%%%%%%%%%
\subsection{The BPS and almost-BPS limits of solutions with an Israel-Wilson base}
%%%%%%%%%%%%%%%%%%%%%%

\begin{figure}[t]
 \centering
    \includegraphics[width=10cm]{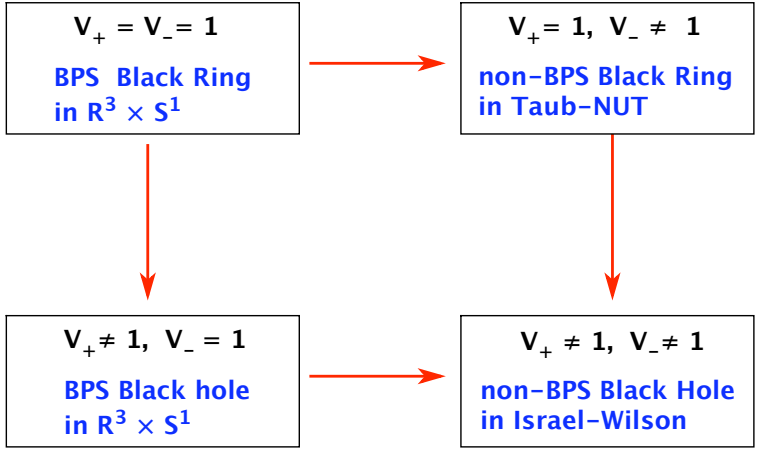}
    \caption{\it \small This diagram represents four classes of
      solutions that can be obtained from our solution for
      various values of the Israel-Wilson harmonic functions. When
      both $V_+$ and $V_-$ are constant, the solution describes a BPS
      black string in $\IR^3\times S^1$. Turning on a
      $\overline{{\it{KKm}}}$ charge at the center of the space
      ($V_- \neq 1$) the space becomes Taub-NUT, the black ring is
      non-BPS and the solution belongs to the almost-BPS Ansatz.
      Turning on a $\it{KKm}$ charge at the location of the ring,
      we obtain a BPS D6-D4-D2-D0 black hole. Turning on both types of
      KKm charges ($V_+ \neq 1$,$V_- \neq 1$), we obtain the more
      general non-BPS solution constructed here: a D6-D4-D2-D0 BPS
      four-charge black hole in a $\overline{\mathrm{D6}}$
      background.}
\label{fig1}
\end{figure}

The solution presented here seems to be somewhat complicated, but its
physical interpretation is rather straightforward. Take first $V_+$ to
be $1$ (by setting the parameter $\alpha$ to zero). As we already
remarked, the metric then becomes the usual Taub-NUT metric, with
negative orientation. Looking at the complete solution, we see that it
becomes the non-BPS black ring in Taub-NUT solution found in
\cite{Bena:2009ev}\footnote{In \cite{Bena:2009ev}, the solution is
  written in the $\varepsilon=-1$ convention.}. From a
four-dimensional perspective this is a two-center solution where one
center is a D4-D2-D0 black hole located at $z=R$ and the other is a
pure $\overline{\mathrm{D}6}$ brane located at $r=0$. Despite the fact
that both objects are locally-BPS, the relative orientation of the
$\overline{\mathrm{D}6}$ and the three D2 branes (which determine,
locally, the Killing spinors of the D4-D2-D0 black hole) makes the
full configuration non-BPS.

On the other hand, setting $Q_{\bar{6}}$ to zero, and hence $V_-=1$,
one can see that the solution becomes BPS and describes a
four-dimensional black hole with D6, D4, D2 and D0 charges. The
singular part in the D4 harmonic function can be traded, via a gauge
transformation \cite{Bena:2005va, Bena:2008wt} for a non-trivial
Wilson line at infinity, and thus this black hole is in fact a BMPV
black hole located at the tip of Taub-NUT (which is now located at at
$z=R$ because we set $Q_{\bar{6}} = 0$), or a single-center D6-D2-D0
black hole from a four-dimensional perspective. For this solution the
relative orientation of the D6 and D2 branes match, and the solution
preserves four supercharges. These two limits are summarized in the
Figure \ref{fig1}. If one now takes both the D6 and the
$\overline{\mathrm{D}6}$ charges to zero, the solution becomes a BPS
D4-D2-D0 four-dimensional black hole, which lifts to a BPS
three-charge three-dipole-charge black string in $\IR^{3,1}\times S^1$
\cite{splittinghairs}.

Having taken these limits, it is now clear that the general solution
with an Israel-Wilson base describes a two-center configuration, where
one of the centers has D6,D4,D2, and D0 charges and is locally-BPS,
and the other has $\overline{\mathrm{D}6}$ charge. Of course, an Israel-Wilson
solution with multiple D6 branes of opposite orientations is only
possible when other charges and fluxes are turned on (\ref{EinVsimp}).
Indeed, the D6 and $\overline{\mathrm{D}6}$ charges attract each other
and in the absence of other branes, there is nothing to balance this
attraction. Introducing D4, D2 and D0 branes creates new interactions:
the D4 branes are also attracted, the D2's feel no force, and the D0's
are repelled, and thus balance becomes possible.

Note that upon flipping the sign of $\varepsilon$ one could also
obtain a solution where the D6 charge becomes anti-D6 charge and
vice versa; this solution should describe an intrinsically non-BPS
$\overline{\mathrm{D}6}$-D2-D0 black hole in a background of a D6
brane that is mutually-BPS with respect to the three sets of D2
branes. When the $\overline{\mathrm{D}6}$ charge is zero the solution
should become a BPS black ring in Taub-NUT \cite{ringTaubNutBPS} and
when the D6 charge is zero it should becomes the almost-BPS
non-rotating $\overline{\mathrm{D}6}$-D2-D0 black hole
\cite{nonrotating-BH}. When both the $\overline{\mathrm{D}6}$ and the
D6 charges are zero, this solution should reduce again to the D4-D2-D0
BPS black hole whose five-dimensional lift is the M5-M2-P black string
(or the infinite black ring) of \cite{splittinghairs}.

%%%%%%%%%%%%%%%%%%%%%%%%%%%%%%%%%%%%%%
\section{Spectral Flow and the Israel-Wilson metric}
%%%%%%%%%%%%%%%%%%%%%%%%%%%%%%%%%%%%%%

In this section we study the action of a spectral flow
transformation \cite{Bena:2008wt} on the solution describing an
almost-BPS black ring in Taub-NUT \cite{Bena:2009ev}, and show that it
yields the $\varepsilon=1$ solution with the Israel-Wilson base found
in the previous section. We also argue that all the solutions that are
constructed starting from a Euclidean electrovac solution given by the
Israel-Wilson metric can be generated by the spectral flow of a
more-standard ``almost-BPS'' solution.

%%%%%%%%%%%%%%%%%%%%%%%%%%%%%%%%%%%%%%%%%%
\subsection{The D1-D5-P duality frame}
%%%%%%%%%%%%%%%%%%%%%%%%%%%%%%%%%%%%%%%%%%

The spectral flow transformation presented in \cite{Bena:2008wt} is a
very useful tool for generating new asymptotically $\IR^{3,1}\times
S^1$ solutions of five-dimensional $U(1)^3$ ungauged supergravity (or
of the STU model in four dimensions) by starting from other such
solutions. In asymptotically $AdS_3 \times S^3$ spaces this
transformation is the gravity counterpart of a symmetry of the dual
CFT, and it is most naturally performed upon dualizing the solution to
the D1-D5-P duality frame \cite{spec-flow-AdS,spec-flow-samir}. In this frame the
solution, which is invariant along the four internal directions
wrapped by the D5 branes, corresponds to a solution of six-dimensional
ungauged supergravity \cite{gmr,oisin-mac}.   The spatial section of
the metric can be written as a $T^2$ fibration over an $\IR^3$ base,
where the $T^2$ is made up by the fiber of the Taub-NUT base space and
by the (internal) direction common to the D1 and the D5 branes.
Spectral flows can then be recast as simply a subgroup of the  group, $SL(2,\ZZ)$, of global
diffeomorphisms on this $T^2$. Thus, from a
six-dimensional point of view, spectral flow is just a change of
coordinates, mixing two different $U(1)$'s. However, upon dualizing
back to the duality frame where the charges correspond to three sets
of M2 branes, the resulting solution, which is again a solution of
$U(1)^3$ supergravity, differs rather non-trivially from the original
one\footnote{Note that to go from a six-dimensional supergravity
  solution to the final solution of five-dimensional supergravity one
  does not KK reduce  the six-dimensional solution; rather one
  trivially uplifts it to a solution of type IIB supergravity,
  performs three T-dualities, then uplifts the resulting solution to
  M-theory, and then reads off the new solution of five-dimensional
  ungauged supergravity}.

\medskip

To perform a spectral flow we need to find the metric and RR
gauge-field of the solution dualized to the D1-D5-P duality frame.
This can be achieved starting from the eleven-dimensional supergravity
solution, dimensionally-reducing along one of the torus directions,
and then performing three T-dualities, as explained in detail in
\cite{Bena:2008dw}. This yields
\bea
ds^2&=& -{1\over \sqrt{Z_1 Z_2} Z_3}(dt+k)^2 + {Z_3\over \sqrt{Z_1 Z_2}} (dy+A^3)^2 +
\sqrt{Z_1 Z_2}ds^2_4 + \sqrt{Z_2\over Z_1} \sum_{a=1}^4 dx_a^2 \,, \\
C^{(2)}&=& A^1\wedge (dy+A^3)+B^{(1)}\wedge {dt+k\over Z_3}+\gamma_2\,,
\label{d1d5p}
\eea
where
\be
A^I=B^{(I)}-{dt+k\over Z_I} \,,
\ee
and
\be
d\gamma_2 = *_4 d Z_2 - B^{(1)} \wedge \Theta^{(3)}\,.
\label{gammaeq}
\ee
For convenience we take $\varepsilon=1$; the result for the other sign is equally straightforward to obtain.

%%%%%%%%%%%%%%%%%%%%%%%%%%%%%
\subsection{The action of spectral flow}
\label{SpecFlowSection}
%%%%%%%%%%%%%%%%%%%%%%%%%%%%%

We start from the solution of the ``almost BPS'' equation presented in
\cite{Bena:2009ev}, corresponding to a non-BPS black ring\footnote{The
  solution in \cite{Bena:2009ev} was first written in the $\varepsilon=-1$ convention, ie $V=V_+$, and
  not the $\varepsilon=+1$ one we use here. One can rewrite the
  solution in the new convention by taking $V=V_-$, and changing the
  signs of the base-space vectors. }
and thus we assume that the base metric $ds^2_4$ has GH form:
\be
ds^2_4 = V_-^{-1}(d\psi+\vec{A}.d\vec{y})^2 + V_- ds^2_3\,,\quad \vec{\nabla}\times\vec{A} = -\vec{\nabla}V_-\,.
\label{VminusGH}
\ee
The one-form potentials are:
\be
B^{(I)} = K_I (d\psi+{A})+{b}_I\,,\quad k = \mu (d\psi+{A}) + {\omega}\,,
\label{OppStubes}
\ee
where $I=1,2,3$, the $K_I$ are harmonic and the ${b}_I$ satisfy the equation:
\be
\vec{\nabla} \times {\vec{b}}_I = -V_- \vec{\nabla} K_I + K_I \vec{\nabla} V_-\,.
\ee
In order to perform the spectral flow, we also need to decompose the two-form, $\gamma_2$, as
\be
\gamma_2= (d\psi+A)\wedge \gamma_2^{(\psi)}+\gamma_2^{(b)}\,,
\ee
where $\gamma_2^{(b)}$ is a two form on the three-dimensional space defined by $ds^2_3$.

Note that the equation for $Z_2$,
\be
d*_4 d Z_2 = \Theta^{(1)}\wedge \Theta^{(3)}  \,,
\ee
implies
\bea
\nabla^2 Z_2 &=& V_- \nabla^2 (K_1 K_3) = \vec{\nabla}.(V_- \vec{\nabla} (K_1 K_3))-\vec{\nabla} V_-.\vec{\nabla} (K_1 K_3)\nn
&=& \vec{\nabla}.(V_- \vec{\nabla}(K_1 K_3) + \vec{A} \times \vec{\nabla} (K_1 K_3))
\eea
and hence
\be
\vec{\nabla} Z_2 = V_- \vec{\nabla} (K_1 K_3) + \vec{A} \times \vec{\nabla} (K_1 K_3)+ \vec{\nabla} L_2\,.
\ee
The equation satisfied by $\gamma_2$
\be
\vec{\nabla}\times \vec{\gamma}_2^{(\psi)} = \vec{\nabla} Z_2 - V_- K_1 \vec{\nabla}K_3 + \vec{b}_1 \times \vec{\nabla}K_3 = \vec{\nabla} L_2 - \vec{\nabla} \times (K_3 \vec{b}_1 + K_1 K_3 \vec{A})\,,
\ee
implies
\be
\gamma_2^{(\psi)} = -K_3 b_1 -  K_1 K_3 A + \hat{\gamma}_2^{(\psi)}\,\,\mathrm{with}\,\, \vec{\nabla} \times \vec{\hat{\gamma}}^{(\psi)}_2 = \vec{\nabla} L_2\,.
\label{gamma21}
\ee
Similarly one can define a two-form
\be
\gamma_1= (d\psi+A)\wedge \gamma_1^{(\psi)}+\gamma_1^{(b)}\,,
\ee
that satisfies
\be
d\gamma_1 = *_4 d Z_1 - B^{(2)} \wedge \Theta^{(3)}\,.
\ee
One has
\be
\vec{\nabla} \times \vec{\gamma}_1^{(\psi)} = \vec{\nabla} Z_1 - V_- K_2 \vec{\nabla} K_3 + b_2 \times \vec{\nabla} K_3 = \vec{\nabla} L_1 - \vec{\nabla} \times (K_3 \vec{b}_2 + K_2 K_3 \vec{A})
\ee
which implies
\be
\gamma_1^{(\psi)} = - K_3 b_2 - K_2 K_3 A + \hat{\gamma}_1^{(\psi)}\,\,\mathrm{with}\,\, \vec{\nabla} \times \vec{\hat{\gamma}}^{(\psi)}_1 = \vec{\nabla} L_1\,.
\label{gamma11}
\ee

Spectral flow mixes the internal $U(1)$ coordinate $y$, associated with the momentum charge, with the GH fiber, $\psi$. Explicitly, this is just the change of coordinates
\be
\psi\to \psi+\alpha \,y\,.
\label{sfcoord}
\ee
To find the transformation of the metric coefficients, one performs the change of coordinates (\ref{sfcoord}) and rewrites the metric and gauge field in the exact same form as (\ref{d1d5p}). Defining the harmonic function $V_+$ by
\bea \label{UK3}
 V_+=1+\alpha K_3,
\eea
the transformed metric is
\be
ds^2_4 ~=~ (V_+V_-)^{-1}(d\psi+\vec{\widetilde{A}}.d\vec{y})^2 + V_+V_- ds^2_3\,,  \qquad
\widetilde{A} ~=~ A-\alpha b_3.
\ee
Note that $\widetilde{A}$ now satisfies:
\be
 \vec{\nabla}\times \vec{\widetilde{A}} = V_-\vec{\nabla}V_+ - V_+ \vec{\nabla}V_-.
\ee
The rest of the fields can be recast in the exact same form as before, with the new coefficients (obtained after a fair amount of of algebra) given by:
\bea
&&\widetilde K_1=K_1 - \alpha \, {Z_2\over V_+V_- }\,,\quad \widetilde K_2=K_2 - \alpha \, {Z_1\over V_+V_- }\,,\quad \widetilde K_3 = {K_3\over V_+}\,, \\
&&\widetilde b_1= V_+ \, b_1 + \alpha \, \gamma^{(\psi)}_2\,,  \quad \widetilde b_2= V_+ \, b_2 + \alpha \, \gamma^{(\psi)}_1\,, \quad {\widetilde b}_3=b_3 \\
&&\widetilde Z_1 = {Z_1\over V_+}\,,\quad \widetilde Z_2 = {Z_2\over V_+}\,,\quad \widetilde Z_3 = V_+ Z_3 - 2 \, \alpha \, \mu + \alpha^2 {Z_1 Z_2 \over V_+V_-}\\
&&\widetilde \mu ={1\over V_+}\Bigl(\mu-\alpha \, {Z_1 Z_2\over V_+V_-}\Bigr)\,,\quad \widetilde\omega=\omega\,.
\label{sf}
\eea

This is exactly the solution with an Israel-Wilson base constructed in
Section \ref{IWmetrics}.  In particular, the relation (\ref{UK3}), which is the same as (\ref{relaKV2}), between the harmonic function
$V_+$ corresponding to the D6 charge and one of the harmonic functions
corresponding to D4 charge emerges directly from the spectral flow transformation.

While this approach to obtaining solutions is rather different from
the one outlined in Section \ref{IWmetrics} in that it does not
involve starting from a non-trivial Einstein-Maxwell electrovac
solution but from a Ricci-flat metric, the resulting solution is the
same. This greatly simplifies the regularity analysis, as we know that
spectral flow always transforms regular solutions into regular
solutions. Hence the regularity of the D6-D4-D2-D0 black hole is
ensured by the regularity of the non-BPS black ring in Taub-NUT, which
yields the regularity conditions outlined in the previous section
(\ref{regularityring}).

%%%%%%%%%%%%%%%%%%%%%%%%%%%%%
\subsection{Spectral flow and  smooth horizonless multi-center solutions.}
%%%%%%%%%%%%%%%%%%%%%%%%%%%%%

One of the driving forces in our effort to construct large classes of
multi-center non-BPS solutions is to obtain smooth horizonless
solutions that have the same charges and mass as non-BPS black holes
with a macroscopically-large horizon area. For BPS black holes, the
existence of large classes of such solutions brings considerable
support to the fact that these black holes should be thought of
statistical ensembles of horizonless configurations, thus realizing
the fuzzball proposal (see \cite{fuzzball-reviews} for reviews) for this class of black holes. We
would like to extend this to non-BPS black holes.

The most obvious way to look for such non-BPS multi-center
horizonless solutions is to use the almost-BPS Ansatz. However, in this
Ansatz the anti-self-dual two-forms that one can turn on (for example
the harmonic forms dual to the the two-cycles of a multi-center Taub-NUT space)  source strongly
singular solutions to the equations of motion. Hence, at least at first glance, no smooth horizonless
solutions exist.

The next obvious place to search for such solutions is in the
floating-brane Ansatz. One obvious way to do this is to construct
solutions explicitly when the base space is has an Israel-Wilson metric. For
particular values of the D4, D2 and D0 charges, the D6 brane of the
two-center floating-brane solution we constructed in Section 4.4 can
become a 16-supercharge fluxed D6 brane, and the five-dimensional lift
of this D6-$\overline{\rm D6}$ solution is completely smooth. The
Israel-Wilson base space has one two-cycle running between the pole of
$V_+$ and the pole of $V_-$, and the non-trivial flux on this two-cycle
is responsible for keeping the D6 and the $\overline{\rm D6}$ apart,
much like for BPS solutions.

Another way to obtain smooth horizonless solutions is to use spectral
flow.  It is well known that in the appropriate IIB frame a two-charge
supertube with D1 and D5 charges corresponds to a completely regular
geometry.  Furthermore, using spectral flow, we can change coordinates
and then dualize a BPS solution containing such a supertube in a
multi-Taub-NUT space into a completely regular multi-Taub-NUT
five-dimensional solution with fluxes supported on bubbles
\cite{Bena:2008wt}.  On the other hand, a solution with multiple
supertubes of different types (with different dipole charges) cannot
simultaneously be dualized via one spectral flow to a smooth geometry.
This needs to be done by three subsequent spectral flows, which
transform every type of supertube into a Taub-NUT center. Since the
near-tube geometry is the same in a BPS and in an almost-BPS solution,
we expect the spectral flow to transform multiple supertubes
in an almost-BPS solution into a smooth non-BPS horizonless solution with
multiple distinct fluxes supported on bubbles.

To illustrate this, consider a single supertube in a Taub-NUT geometry of ``opposite
orientation.''  That is, the base space is of the form
(\ref{VminusGH}) while the supertube magnetic dipoles are given by
(\ref{OppStubes}).  If $K_1= K_2 = 0$ this supertube has only one
dipole charge, and it can be arranged to give a completely regular
geometry.  However, as explained in \cite{Bena:2009ev}, even if this
solution is written as an almost-BPS solution, it still preserves four
supersymmetries\footnote{Essentially because the supertube only has
  two D2 charges, that are mutually-BPS with respect to a D6 brane
  irrespective of its orientation.}. One can now perform a spectral
flow on this solution exactly as in Section \ref{SpecFlowSection} and
obtain a floating-brane solution with an Israel-Wilson base that has
$V_-$ unchanged and $V_+$ given by (\ref{UK3}).  The spectral flow
transformation preserves the regularity of the solution and replaces
the supertube by a fluxed ${\mathrm{D}6}$ brane, which is also
perfectly regular. Hence, one obtains the smooth D6-$\overline{\mathrm
  D6}$ solution with non-trivial flux described above\footnote{It is
  worth commenting on how much supersymmetry this solution preserves.
  On one hand, we have obtained this solution by spectral flow from a
  supersymmetric solution. Since spectral flow is a combination of
  coordinate transformations and dualities, one would expect the
  resulting solution to still be supersymmetric. On the other hand,
  the resulting solution has an Israel-Wilson base, and, as proved in
  \cite{5dsugra,gutowski-reall}, such solutions should not be
  supersymmetric, since all supersymmetric solutions must have a
  Hyper-K\"ahler base. The resolution of the puzzle is a generalization
  of that described in \cite{Dunajski:2006vs}:  For these very special solutions the warp factors and angular
  momentum vector are such that if one makes a coordinate
  transformation of the type $\psi \rightarrow \psi + \alpha t$, and
  rewrites the metric as a time fibration over a four-dimensional
  base, this base space can be made hyper-K\"ahler. Hence this particular
  floating-brane solution is secretly BPS.}.

One can take this procedure further, and consider two or three
different types of supertube in a GH geometry of the opposite
orientation. Unlike the single supertube, this solution is no longer
BPS, as the holonomy of the base metric is inconsistent with the
supersymmetry projections associated with all the supertubes (the
solution has three D2 and one $\overline{\mathrm{D}6}$ charge).

If one now makes several spectral flows to convert each species of
supertube to fluxes supported by geometry, the result must be regular
for exactly the same reason that the BPS supertubes produce regular
geometries after spectral flow: The almost-BPS supertubes are locally
identical to BPS supertubes and so the spectral flow cannot generate
singularities.  The result of such a multiple spectral flow must
therefore be a completely regular, non-BPS geometry with fluxes in
five dimensions.  We expect that these solutions will go well beyond the Israel-Wilson
class:  Indeed, the metric coefficients of the base will generically involve products of
more than two functions.   We also expect this method to yield large classes of
smooth horizonless non-BPS scaling solutions, which will be
instrumental in extending the fuzball proposal to non-BPS extremal
black holes.

%%%%%%%%%%%%%%%%%%%%%%%%%%%%%%%%%%%%%
\section{Conclusion}
%%%%%%%%%%%%%%%%%%%%%%%%%%%%%%%%%%%%%

We have solved the equations of motion for five-dimensional ungauged
supergravity coupled to three $U(1)$ gauge fields using a
floating-brane Ansatz in which M2 branes feel no force, and hence the
warp factors and the Maxwell electric potentials are equal. Upon
making a simplifying assumption we have obtained a new class of
non-BPS solutions, that are constructed starting not from a
hyper-K\"ahler base (like the BPS and almost-BPS solutions) but from a
much more general Einstein-Maxwell Euclidean electrovac solution, and
solving a new linear system of equations for the warp factors and
magnetic potentials. These ``simplified floating brane'' solutions are
much more general than both the BPS and almost-BPS solutions, and
reduce to them when the electrovac Maxwell fields are self- or
anti-self-dual, and the base becomes Ricci-flat. We have also noted
that this implies that the BPS and ``almost-BPS'' equations yield full
solutions of the supergravity equations of motion not only when the
base is hyper-K\"ahler (as previously thought) but also when the base
is Ricci-flat. A few such solutions are presented in
\cite{Bena:2009qv}.

The floating brane Ansatz requires that the warp factors and the
electric potentials are equal.  Not only does this result in no force
upon appropriate brane probes but it also means that the mass of our
solutions will be linear in their M2 brane charges.  Hence, the
non-BPS solutions that result from our Ansatz will naturally describe
single or multiple extremal black holes, as well as smooth horizonless
solutions that have the same charges and mass as extremal black holes.
One can think about these solutions as having D-brane components that
locally preserve some supersymmetries but whose supersymmetries are
either globally incompatible with one another, or are broken by the
gravitational background. Thus, we cannot hope to use this approach to
obtain completely general non-extremal solutions, but only
interesting sub-classes (like the Running-Bolt \cite{Bena:2009qv}) in
which the mass is linear in the electric charges.

We have illustrated our method by finding a new two-center solution
that has as a base space an Israel-Wilson metric, and that describes a
D6-D4-D2-D0 black hole in the background of an anti-D6 brane. We have
also shown that spectral flow can be used to map some of the new
solutions into previously-known ``almost-BPS'' solutions.

While we have presented the material here in what seems to us as the
natural expository order, we initially discovered examples of
Israel-Wilson solutions by considering spectral flows of non-BPS
solutions on GH spaces. This closely parallels the history of the
discovery of the importance of ambi-polar base spaces for the
constructions of BPS horizonless bubbling solutions; the crucial first
examples of such metrics were obtained by spectral flow in
\cite{spec-flow-samir,Giusto:2004kj} and then greatly generalized and more deeply
understood via geometric transitions in 
\cite{Bena:2005va,Berglund:2005vb}. It is thus evident that spectral flow is a very
powerful tool in suggesting completely new classes of
physically-interesting solutions that can then be further generalized.
In this paper we have only really exploited a single spectral flow and
it is therefore very natural to continue exploring the
more-complicated solutions generated by two or three spectral flows.
It would be very interesting to see if one could find an Ansatz
(similar to the floating-brane Ansatz used here) that describes
such solutions, and if they could still be obtained using a linear
procedure.

We have also obtained a smooth two-center non-BPS solution that
describes a D6 and an anti-D6 brane kept in equilibrium by flux on the
two-cycle between them. We would like to note that there exists a
solution where such branes are kept in equilibrium by a background
magnetic field \cite{gross-perry,Sen:1997pr}.  Furthermore, the
non-BPS running Kerr-Taub-Bolt solutions recently constructed in
\cite{Bena:2009qv} can be thought of as describing a D6 and an anti-D6
brane kept in equilibrium by {\it both } flux on the bolt and
background magnetic field. It is quite likely that the floating-brane
solutions constructed using Euclidean Reissner-Nordstrom electrovac
base spaces \cite{ReissnerNord} will have a similar interpretation. It
would be interesting to explore all the possible ways of constructing
a non-BPS D6 - anti D6 system in equilibrium, and to see whether one
might be able to build two distinct supergravity solutions with the
same brane interpretation.

We have made some conjectures as to the form of the solutions that
will arise from multiple spectral flows, and they certainly will go
beyond the Israel-Wilson electrovac backgrounds.  We also expect
that such solutions will describe even more general black holes and
horizonless solutions.  On the other hand, putting these solutions in
the six-dimensional form that makes spectral flow into a mere
coordinate transformation makes it clear that a spectral flow will
always preserve the same flat $\IR^3$ section of the base.  Spectral
flow simply does not affect this $\IR^3$ structure and so even if the
solutions that come out after multiple spectral flows are more
general, the spatial metric will still have the form of $U(1)^2$
fibrations over an $\IR^3$ base.  At first glance it appears that
these solutions will not be general enough to describe {\it all}
extremal black holes and black rings. For example, neither the
extremal overspinning Rasheed-Larsen black hole \cite{rasheed-larsen}
nor the extremal non-BPS three-charge black ring
\cite{Elvang-Emparan-extremal-ring} can be written as fibrations over
a flat $\IR^3$ base. However, it is likely they will describe very
large and non-trivial classes of non-BPS solutions, that will yield
interesting physics.

%%%%%%%%%%%%%%%%%%%%%%%%%%%%%%%%%%%%%
\bigskip
\leftline{\bf Acknowledgments}
\smallskip
We would like to thank N. Bobev, K. Goldstein, S. Katmadas and G.
Gibbons for valuable discussions.  NPW is grateful to the IPhT(SPhT),
CEA-Saclay for hospitality while this work was done. The work of IB,
CR and SG was supported in part by the DSM CEA-Saclay, by the ANR
grants BLAN 06-3-137168 and 08-JCJC-0001-01, and by the Marie Curie IRG
046430. The work of NPW was supported in part by DOE grant
DE-FG03-84ER-40168.

%%%%%%%%%%%%%%%%%%%%%%%%%%%%%%%%%%%%%

%%%%%%%%%%%%%%%%%%%%%%%%%%%%%%%%%%%%%

\end{document}